\documentclass[conference]{IEEEtran}
\IEEEoverridecommandlockouts
\usepackage[noadjust]{cite}
\usepackage{amsmath,amssymb,amsfonts}
\usepackage{algorithmic}
\usepackage{graphicx}
\usepackage{textcomp}
\usepackage{flushend}
\usepackage{bbm}
\usepackage{dsfont}
\usepackage[caption=false,font=scriptsize]{subfig}
\usepackage{xcolor}
\def\BibTeX{{\rm B\kern-.05em{\sc i\kern-.025em b}\kern-.08em
    T\kern-.1667em\lower.7ex\hbox{E}\kern-.125emX}}

\newtheorem{identity}{Identity}
\newtheorem{definition}{Definition}

\begin{document}

\title{Zak-OTFS for Mutually Unbiased  \\ Sensing and Communication\\
\thanks{
The Duke team is supported by the NSF under grants 2342690 and 2148212, in part by funds from federal agency and industry partners as specified in the RINGS program, and in part by the Air Force Office of Scientific Research under grants FA 8750-20-2-0504 and FA 9550-23-1-0249. \\
$*$ denotes equal contribution. The author order was decided by a coin toss. \\
This work may be submitted to the IEEE for possible publication. Copyright may be transferred without notice, after which this version may no longer be accessible.}
}

\author{\IEEEauthorblockN{Nishant Mehrotra$^*$}
\IEEEauthorblockA{\textit{Electrical and Computer Engineering} \\
\textit{Duke University}\\
Durham, USA \\
nishant.mehrotra@duke.edu\vspace{-7mm}}
\and
\IEEEauthorblockN{Sandesh Rao Mattu$^*$}
\IEEEauthorblockA{\textit{Electrical and Computer Engineering} \\
\textit{Duke University}\\
Durham, USA \\
sandesh.mattu@duke.edu\vspace{-7mm}}
\and
\IEEEauthorblockN{Robert Calderbank}
\IEEEauthorblockA{\textit{Electrical and Computer Engineering} \\
\textit{Duke University}\\
Durham, USA \\
robert.calderbank@duke.edu\vspace{-7mm}}
}

\maketitle
\begin{abstract}
Waveforms with ideal ambiguity functions are fundamental to integrated sensing and communication, to active sensing (radar), and to uplink multiple access. We describe a general method of constructing waveforms using the discrete Zak transform (DZT) to convert sequences of length $MN$ in the time domain to waveforms in the delay-Doppler (DD) domain, each of which is defined by an $M\times N$ quasi-periodic array. The DZT preserves inner products, and we show that phase coded waveforms used in radar (CAZAC sequences) determine noise-like waveforms in the DD domain, each with low Peak to Average Power Ratio. In a Zak-OTFS communication system, we show that these waveforms are mutually unbiased with respect to every carrier and use them to integrate sensing and communication as spread pilots. We view each waveform as a linear combination of Zak-OTFS carriers and show that the self-ambiguity function is supported on a discrete line in the integers modulo $MN$. The sidelobes are significantly lower than the original CAZAC sequence, and the advantage of discrete support is better localization/resolution in delay and Doppler compared with standard methods based on chirps or tones. We show that the absolute value of the cross-ambiguity function for pairs of waveforms in the same family is small and constant. This property makes the waveforms ideal preambles in the 2-step RACH protocol introduced in Release 15, 3GPP to enable grant-free multiple access. The characteristics of the cross-ambiguity function make it possible to simultaneously detect multiple preambles in the presence of mobility and delay spread.
\end{abstract}


\section{Introduction}
\label{sec:intro}
In 1953, Philip Woodward \cite{woodward2014probability} suggested that we view a radar scene as an unknown operator parameterized by delay and Doppler, and that we view radar waveforms as questions that we ask the operator. What makes a good question is lack of ambiguity in the answer. 

Matched filtering correlates the radar return with the transmitted waveform, making it possible to identify targets by the path length (or, by the path delay), and by the Doppler shift determined by the radial velocity. The radar ambiguity function \cite{woodward2014probability}, \cite{auslander1985radar}, \cite{moran2001mathematics} is the point-spread function for the delay-Doppler (DD) plane, and it captures the way matched filtering blurs the radar scene. Radar engineers construct waveform libraries for tracking applications consisting of a fixed waveform that has been linearly frequency modulated or \textit{chirped} at several rates (see \cite{howard2004waveform} for more details). Chirping rotates the ambiguity function. A target that might not be visible because of the blur associated with the initial waveform might become visible when the blur is rotated. 

Given a continuous-time waveform, it is possible to shape the ambiguity function by using a sequence of complex phases to modulate the waveform (see \cite{benedetto_phasecoded} for a comprehensive survey). The characteristics of the ambiguity function depend on the correlation properties of the sequence of phases \emph{and} the choice of initial waveform. The survey \cite{benedetto_phasecoded} only considers the discrete ambiguity function which records the influence of the correlation properties of the sequence of phases. 

We depart from traditional phase coding by using a sequence of complex phases to directly specify a continuous-time waveform. We start from the orthonormal basis of Zak-OTFS carrier waveforms, and we use a sequence of complex phases to specify a linear combination of the basis waveforms with ideal ambiguity properties. The Zak-OTFS carrier waveforms are pulses in the DD domain regularly spaced on an $M\times N$ period grid. They are quasi-periodic functions with delay period $M$ and Doppler period $N$. Section \ref{sec:sysmodel} introduces the discrete Zak transform (DZT) which preserves inner products as it maps sequences in the time domain (TD) with period $MN$ to quasi-periodic $M\times N$ arrays in the DD domain. Section \ref{sec:ambiguityfunc} introduces discrete ambiguity functions in the TD and the DD domain, and Section \ref{sec:cazac} shows that these notions are equivalent. It follows that we can start with Constant Amplitude Zero Autocorrelation (CAZAC) sequences with ideal ambiguity properties in the TD and use the DZT to transfer these properties to the DD domain. Section \ref{sec:cazac} considers families of CAZAC sequences of length $MN$ where the $n$th term is a fixed $MN$th root of unity raised to a power which is a quadratic function of $n$. We relate the form of the quadratic exponent to the support of the ambiguity function and prove that it is a line modulo $MN$. For a parallel construction of waveforms using discrete filters in the DD domain where the ambiguity function is supported on a lattice modulo $MN$, see~\cite{Aug2024paper}. The advantage of an ambiguity function supported on a lattice or discrete line is better localization/resolution in delay and Doppler. This is particularly important when a radar needs to detect multiple targets (for a comparison against standard chirp-based waveforms, see~\cite{zakotfs_ltv}). The ambiguity functions in Section~\ref{sec:implications} illustrate the value of designing a continuous waveform rather than a discrete sequence of complex phases.

Each preamble constructed in this paper is mutually unbiased with respect to every OTFS carrier. In a Zak-OTFS communication system, these preambles can serve as spread pilots with a low Peak to Average Power Ratio (PAPR). Pilot and data are superimposed in the same frame, and the effective channel can be read off from the response to the pilot (for details, see~\cite{Aug2024paper}). By minimizing interference between pilot and data we maximize throughput by enabling coexistence of sensing and communication in the same radio resources.

Each family of waveforms constructed in this paper has the property that the inner product of a first waveform in the family with a delay and Doppler shifted version of a second waveform in the family has size $\frac{1}{\sqrt{MN}}$. This property makes it possible to detect multiple preambles in the presence of mobility and delay spread using a receiver with no knowledge of the channel other than the worst case delay and Doppler spreads (for details, see~\cite{preamblepaper}). Our target application is grant-free multiple access in the spirit of~\cite{polyanskiy2017} but in the presence of mobility and delay spread (for details of a wireless uplink aligned with the 2-step RACH protocol in 5G NR, see~\cite{agostini2024evolution}).

\textit{Notation:} $\mathds{1}_{\{\cdot\}}$ denotes the indicator function, and $\delta[\cdot]$ denotes the Kronecker delta function. $x$ denotes a complex scalar, with $x^{\ast}$ denoting its complex conjugate. $\mathbf{x}$ denotes a vector with $n$th entry $\mathbf{x}[n]$, and $\mathbf{X}$ denotes a matrix with $(n,m)$th entry $\mathbf{X}[n,m]$. $\langle \cdot, \cdot \rangle$ denotes the inner product between two vectors. $\lfloor \cdot \rfloor$ denotes the floor function. $\mathbb{Z}$ denotes the set of integers, and $\mathbb{Z}_{N}$ denotes the set of integers modulo $N$. 

\section{Periodic Sequences \& Quasi-Periodic Arrays}
\label{sec:sysmodel}

This Section describes how the discrete Zak transform (DZT) maps periodic sequences in the time domain (TD) to quasi-periodic arrays in the delay-Doppler (DD) domain. Given odd integers $M, N$ we introduce an orthonormal basis for the Hilbert space of $MN$-periodic sequences:
\begin{align}
    \label{eq:timebasis}
    \mathbf{v}_{r,s}[n] = \begin{cases}
        \frac{1}{\sqrt{M}} e^{\frac{j2\pi}{M} sn}, \quad \text{if } rM \leq n < (r+1)M \\ 
        0, \quad \text{otherwise},
    \end{cases}
\end{align}
where $\mathbf{v}_{r,s}[n]$ is repeated periodically with period $MN$. Therefore, the inner product between two sequences can be calculated on any subsequence of length $MN$ as:
\begin{align}
    \big\langle \mathbf{v}_{r,s}, \mathbf{v}_{r',s'} \big\rangle &= \frac{1}{M} \sum_{n = 0}^{MN-1} \mathds{1}_{\{rM \leq n < (r+1)M\}} e^{\frac{j2\pi}{M} sn} \times \nonumber \\ &\hspace{5mm}\mathds{1}_{\{r'M \leq n < (r'+1)M\}} e^{-\frac{j2\pi}{M} s'n} \nonumber \\
    &= \delta[r-r'] \frac{1}{M} \sum_{n = rM}^{(r+1)M-1} e^{\frac{j2\pi}{M} (s-s')n} \nonumber \\
    &= \delta[r-r']  \delta[s-s'],
    \label{eq:timebasis_orth}
\end{align}
where the last expression follows from Identity~\ref{idty:sumrootsofunity}.
\begin{identity}
    \label{idty:sumrootsofunity}
    The sum of all $N$th roots of unity satisfies:
    \begin{align}
        \sum_{n=0}^{N-1}e^{\frac{j2\pi}{N}kn} = \begin{cases}
        N \quad \text{if } \ k \equiv 0 \mod (N) \\
        0 \quad \ \text{otherwise}
        \end{cases}.
    \end{align}
\end{identity}


\begin{definition}[\cite{dzt}]
    \label{def:dzt}
    The discrete Zak transform (DZT) maps an $MN$-periodic sequence $\mathbf{x}[n]$ to a quasi-periodic\footnote{An $M \times N$ array $\mathbf{X}[k,l]$ is quasi-periodic if: $\mathbf{X}[k+nM,l+mN] = e^{j2\pi\frac{nl}{N}} \mathbf{X}[k,l], \text{for all~} n,m \in \mathbb{Z}$.} $M \times N$ array $\mathbf{X}[k,l]$ as follows:
    \begin{align}
        \mathbf{X}[k, l] = \frac{1}{\sqrt{N}} \sum_{p = 0}^{N-1}\mathbf{x}[k+pM]e^{-\frac{j2\pi}{N} pl},
    \label{eq:dzt}
    \end{align}
    where $k,l \in \mathbb{Z}$, and $k$ is the delay and $l$ is the Doppler axis. The $M \times N$ quasi-periodic arrays $\mathbf{X}[k, l]$ form a Hilbert space, and inner products can be calculated on any $M \times N$ subarray. 
\end{definition}

\begin{figure*}[ht]
    \centering
    \includegraphics[clip=true, trim=0in 0in 0in 0.4in, width=0.8\textwidth]{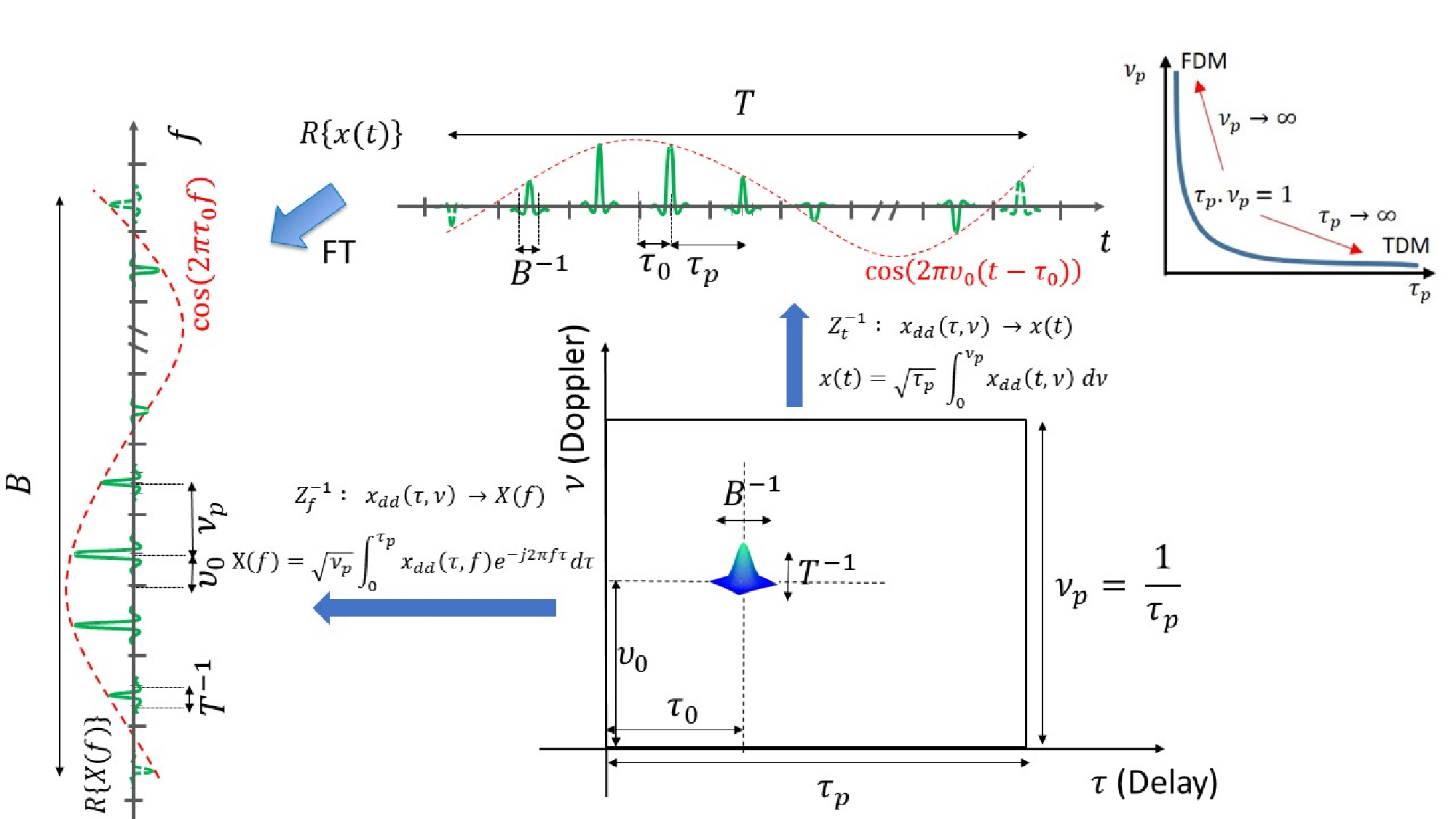}
    \caption{ A DD domain pulse and its TD/FD realizations referred to as TD/FD pulsone. The TD pulsone comprises of a finite duration pulse train modulated by a TD tone. The FD pulsone comprises of a finite bandwidth pulse train modulated by a FD tone. The location of the pulses in the TD/FD pulse train and the frequency of the modulated TD/FD tone is determined by the location of the DD domain pulse $(\tau_0, \nu_0)$. The time duration ($T$) and bandwidth ($B$) of a pulsone are inversely proportional to the characteristic width of the DD domain pulse along the Doppler axis and the delay axis, respectively. The number of non-overlapping DD pulses, each spread over an area $B^{-1}T^{-1}$, inside the fundamental period $\mathcal{D}_0$ (which has unit area) is equal to the time-bandwidth product $BT$ and the corresponding pulsones are orthogonal to one another, rendering OTFS an orthogonal modulation that achieves the Nyquist rate. As the Doppler period $\nu_p\to\infty$, the FD pulsone approaches a single FD pulse which is the FDM carrier. Similarly, as the delay period $\tau_p\to\infty$, the TD pulsone approaches a single TD pulse which is the TDM carrier. Setting $\tau_p\nu_p = 1$, we see that OTFS is a family of modulations parameterized by $\tau_p$ that interpolates between TDM and FDM. The pulses are supported on the \textit{information lattice} $\Delta_{\mathrm{dd}} = \left\{ (k \tau_p/M, l \nu_p/N) \mid k, l \in \mathbb{Z}\right\}.$ }
    \label{fig:pointpulsone} 
\end{figure*}

By Definition~\ref{def:dzt}, the DZT of~\eqref{eq:timebasis} is:
\begin{align}
    \label{eq:timebasis_dzt}
    \mathbf{V}_{r,s}[k,l] &= \frac{1}{\sqrt{N}} \sum_{p = 0}^{N-1}\mathbf{v}_{r,s}[k+pM] e^{-\frac{j2\pi}{N} pl} \nonumber \\
    &= \frac{1}{\sqrt{MN}} \sum_{p = 0}^{N-1} e^{\frac{j2\pi}{M}s(k+pM)} e^{-\frac{j2\pi}{N} pl} \times \nonumber \\ &\qquad \qquad \qquad \quad \mathds{1}_{\{rM \leq k+pM < (r+1)M\}} \nonumber \\
    &= \frac{1}{\sqrt{MN}} \sum_{p = 0}^{N-1} e^{\frac{j2\pi}{M}s(k+pM)} \delta\bigg[p-r+\bigg\lfloor\frac{k}{M}\bigg\rfloor\bigg] \times \nonumber \\ &\qquad \quad \qquad \quad e^{-\frac{j2\pi}{N} pl} \nonumber \\
    &= \frac{1}{\sqrt{MN}} e^{\frac{j2\pi}{M}s(k+(r-\lfloor\frac{k}{M}\rfloor)M)} e^{-\frac{j2\pi}{N} (r-\lfloor\frac{k}{M}\rfloor)l} \nonumber \\ 
    &= \frac{1}{\sqrt{MN}} e^{\frac{j2\pi}{M}sk} e^{-\frac{j2\pi}{N} (r-\lfloor\frac{k}{M}\rfloor)l}.
\end{align}

The DZT preserves inner products since it maps the orthonormal basis $\mathbf{v}_{r,s}$ to the orthonormal basis $\mathbf{V}_{r,s}$.

The Zak-OTFS carrier waveform is a pulse in the DD domain. Fig.~\ref{fig:pointpulsone} illustrates how a quasi-periodic array in the DD domain determines a carrier waveform in the TD which we call a \emph{point pulsone}. Zak-OTFS carriers are parameterized by a delay period $\tau_{p}$ and a Doppler period $\nu_{p}$. We refer the reader to~\cite{bitspaper2,preamblepaper} for details of the system model, including design of transmit \& receive filters. The point pulsone corresponds to a quasi-periodic array $\mathbf{V}_{r,s}$. It is a pulse train modulated by a tone, and this structure can be seen in the basis element $\mathbf{v}_{r,s}$.



\section{Ambiguity Functions}
\label{sec:ambiguityfunc}

We consider discrete sequences that are periodic in time and quasi-periodic in the DD domain. 

\begin{definition}[\cite{benedetto_phasecoded}]
    \label{def:tdamb}
    The \emph{time-domain discrete cross-ambiguity function} of two periodic sequences $\mathbf{x}[n]$ and $\mathbf{y}[n]$, with period $MN$ each, is given by:
    \begin{align}
        \mathbf{A}_{\mathbf{x},\mathbf{y}}[k,l] = \frac{1}{MN}\sum_{n=0}^{MN-1} \mathbf{x}[k+n]\mathbf{y}^{\ast}[n] e^{-\frac{j2\pi}{MN} nl}.
    \end{align}

    When $\mathbf{x}[n] = \mathbf{y}[n]$, $\mathbf{A}_{\mathbf{x},\mathbf{x}}[k, l]$ is referred to as the \emph{time-domain discrete self-ambiguity function} of $\mathbf{x}[n]$.
\end{definition}


\begin{definition}[\cite{bitspaper2}]
    \label{def:ddamb}
    The \emph{DD domain discrete cross-ambiguity function} of two quasi-periodic DD domain signals $\mathbf{X}[k, l]$ and $\mathbf{Y}[k, l]$, with periods $M$ and $N$ respectively along delay and Doppler, is given by:
    \begin{align}
        \mathbf{A}_{\mathbf{X},\mathbf{Y}}[k, l] = \frac{1}{MN}\sum_{k'=0}^{M-1}\sum_{l'=0}^{N-1}&\mathbf{X}[k', l']\mathbf{Y}^{\ast}[k'-k, l'-l] \times \nonumber \\ 
    &e^{-\frac{j2\pi}{MN} (k'-k)l}.
    \end{align}

    When $\mathbf{X}[k, l] = \mathbf{Y}[k, l]$, $\mathbf{A}_{\mathbf{X},\mathbf{X}}[k, l]$ is referred to as the \emph{DD domain discrete self-ambiguity function} of $\mathbf{X}[k, l]$.
\end{definition}


For brevity, in the following sections, we will refer to the time-domain ambiguity function as the TD AF and the DD domain ambiguity function as the DD AF.

\section{Sequences with Desirable DD Ambiguity}
\label{sec:cazac}

\subsection{Equivalence of TD \& DD Ambiguity Functions}
\label{subsec:cazac_af}

In this subsection we show that the ambiguity functions in Definitions~\ref{def:tdamb} \&~\ref{def:ddamb} are equivalent. To this end, we substitute the DZT expression from Definition \ref{def:dzt} into the DD AF from Definition \ref{def:ddamb} and simplify as follows:
\begin{align*}
    \mathbf{A}_{\mathbf{X}, \mathbf{Y}}[k, l] &= \frac{1}{MN^2}\sum_{k'=0}^{M-1}\sum_{l'=0}^{N-1}\sum_{p=0}^{N-1}\sum_{q=0}^{N-1} \mathbf{x}[k'+pM]e^{-\frac{j2\pi}{N}pl'} \times \nonumber\\
    & \hspace{7mm}\mathbf{y}^*[k'-k+qM]e^{\frac{j2\pi}{N}q(l'-l)}e^{-\frac{j2\pi}{MN}l(k'-k)} \nonumber \\
\end{align*}
\begin{align}
    &= \frac{1}{MN^2}\sum_{p=0}^{N-1}\sum_{q=0}^{N-1}\sum_{k'=0}^{M-1}\mathbf{x}[k'+pM] e^{-\frac{j2\pi}{MN}l(k'-k)} \times \nonumber \\
    &\hspace{3mm}\mathbf{y}^*[k'-k+qM]e^{-\frac{j2\pi}{N}ql}\sum_{l'=0}^{N-1}e^{-\frac{j2\pi}{N}l'(p-q)}.
    \label{eq:ddaf_1}
\end{align}

It follows from Identity~\ref{idty:sumrootsofunity} that the inner sum over $l'$ vanishes unless $p=q$, so that:
\begin{align}
    \mathbf{A}_{\mathbf{X}, \mathbf{Y}}[k, l] &= \frac{1}{MN}\sum_{k'=0}^{M-1}\sum_{p=0}^{N-1} \mathbf{x}[k'+pM]\mathbf{y}^*[k'-k+pM] \times \nonumber \\
    &\hspace{5mm}e^{-\frac{j2\pi}{MN}l(k'-k)}e^{-\frac{j2\pi}{N}pl}.
    \label{eq:ddaf_2}
\end{align}

Substituting $t = k'+pM$ and summing over $t$ in \eqref{eq:ddaf_2} gives:
\begin{align}
    \mathbf{A}_{\mathbf{X}, \mathbf{Y}}[k, l] &= \frac{1}{MN}\sum_{t=0}^{MN-1}\mathbf{x}[t]\mathbf{y}^*[t-k]e^{-\frac{j2\pi}{MN}l(t-k)},
    \label{eq:ddaf_3}
\end{align}
which is the TD AF per Definition~\ref{def:tdamb}. 

The equivalence of the TD and DD AFs implies that sequences with desirable TD AF properties must also have desirable DD AF properties. The authors in~\cite{benedetto_phasecoded,benedetto_cazac} show that constant-amplitude zero-autocorrelation (CAZAC) sequences\footnote{For an $MN$-periodic CAZAC sequence $\mathbf{x}[n], n=0, 1, \cdots, MN-1$, CA implies that $\vert\mathbf{x}[n]\vert = 1, \text{for all~} n$ and ZAC implies that the sum $\frac{1}{MN}\sum_{n=0}^{MN-1}\mathbf{x}[n+k]\mathbf{x}^*[n] = \delta[k], \text{for all~} k \in \mathbb{Z}_{MN}$.} are a family of sequences with excellent TD AF properties. Per~\eqref{eq:ddaf_3}, CAZAC sequences must also possess excellent DD AF properties. We thus propose to convert the time-domain CAZAC sequences to the DD domain via the DZT in Definition~\ref{def:dzt}. Since DZT preserves inner products (see Section~\ref{sec:sysmodel}), a CAZAC sequence in time determines a CAZAC sequence in the DD domain.





\subsection{Sequence Construction using CAZACs}
\label{subsec:cazac_seq}

Given odd integers $M,N$, we define a general CAZAC sequence of length $MN$ as follows:
\begin{align}
    \mathbf{x}[n] = e^{\frac{j2\pi}{MN}(\alpha n^2 + \beta n + \gamma)},
    \label{eq:cazac}
\end{align}
where $\alpha, \beta, \gamma \in \mathbb{Z}$, $2\alpha \not\equiv 0 \mod (MN)$. Note that $\mathbf{x}[n]$ is periodic with period $MN$.


The sequence in~\eqref{eq:cazac} is CAZAC since: $(i)$ (CA) $|\mathbf{x}[n]| = 1$, and $(ii)$ (ZAC) $\frac{1}{MN} \sum_{n = 0}^{MN-1} \mathbf{x}[n+k] \mathbf{x}^{\ast}[n] = \frac{1}{MN} \sum_{n = 0}^{MN-1} e^{\frac{j2\pi}{MN}(\alpha (k^2+2nk) + \beta k)}$. Since $2\alpha\not\equiv 0~\mod~(MN)$, it follows from Identity~\ref{idty:sumrootsofunity} that the sum vanishes unless $k \equiv 0 \mod (MN)$.


Special cases of~\eqref{eq:cazac} correspond to \emph{Zadoff-Chu} (ZC) sequences~\cite{zadoff_chu} of root $u$ when $\alpha = \beta = u/2$ and $\gamma = 0$, \emph{Gaussian} sequences~\cite{benedetto_phasecoded} when $\gamma = 0$, and \emph{Wiener} sequences~\cite{benedetto_cazac} when $\beta = \gamma = 0$.

We apply the DZT (Definition~\ref{def:dzt}) to convert the TD sequence to the DD domain:
\begin{align}
    \mathbf{X}[k, l] &= \frac{1}{\sqrt{N}}\sum_{p=0}^{N-1}e^{\frac{j2\pi}{MN}(\alpha(k+pM)^2 + \beta(k+pM) + \gamma)}e^{-\frac{j2\pi}{N}pl} \nonumber \\
    &= \frac{e^{\frac{j2\pi}{MN}\left(\gamma + k\beta\right)}}{\sqrt{N}} \sum_{p=0}^{N-1}e^{\frac{j2\pi}{MN}\left(\alpha(k+pM)^2+pM(\beta-l)\right)}.
    \label{eq:cazac_dd}
\end{align}

Since the DZT preserves inner products (Section~\ref{sec:sysmodel}), we expect the DD array $\mathbf{X}[k, l]$ in~\eqref{eq:cazac_dd} to have excellent DD self- and cross-AF properties.




\section{DD Ambiguity Functions of DD Arrays~\eqref{eq:cazac_dd}}
\label{sec:amb_cazac}

\begin{figure*}
    \subfloat[{ZC Phase Coded TD AF (from \cite{benedetto_phasecoded})}]{\includegraphics[width=0.5\linewidth]{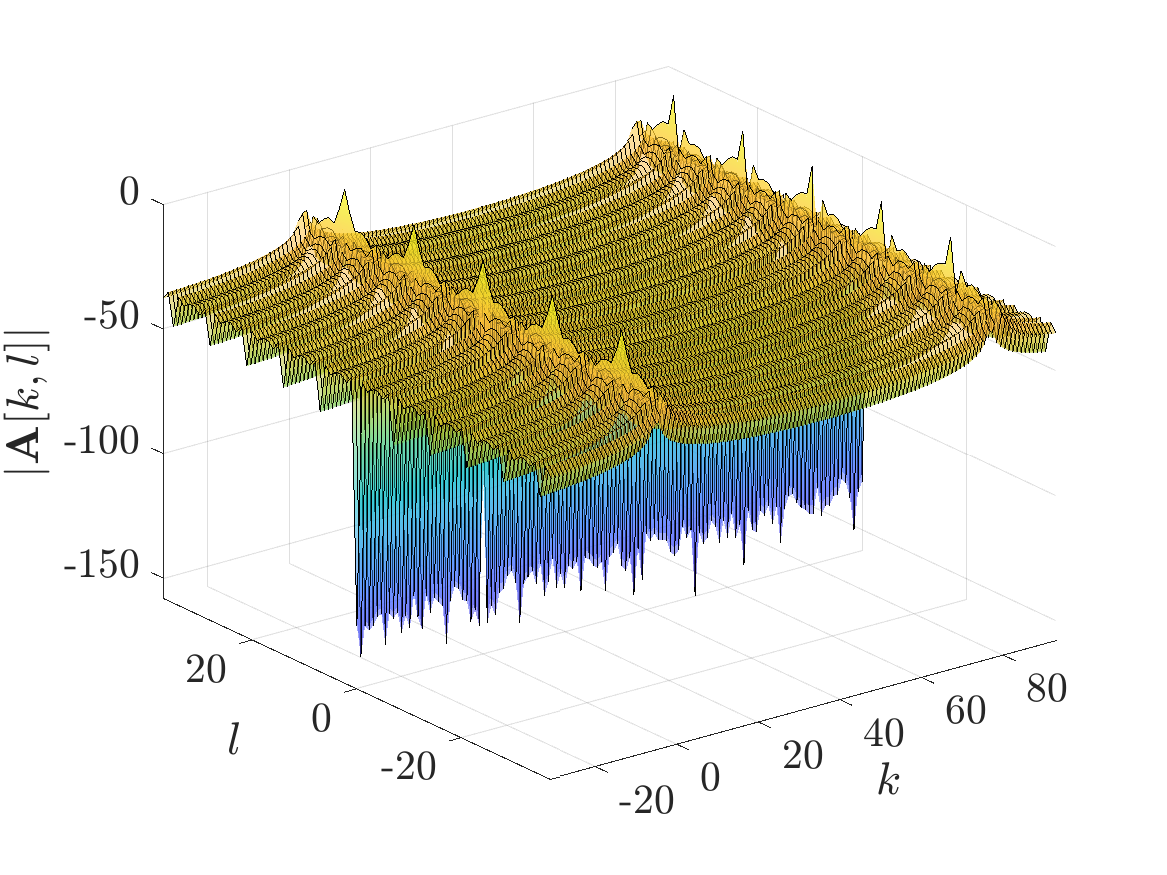}\label{fig:sa_benedetto}}
    \hfill
    \subfloat[{ZC Zak-OTFS DD AF (Proposed)}]{\includegraphics[width=0.5\linewidth]{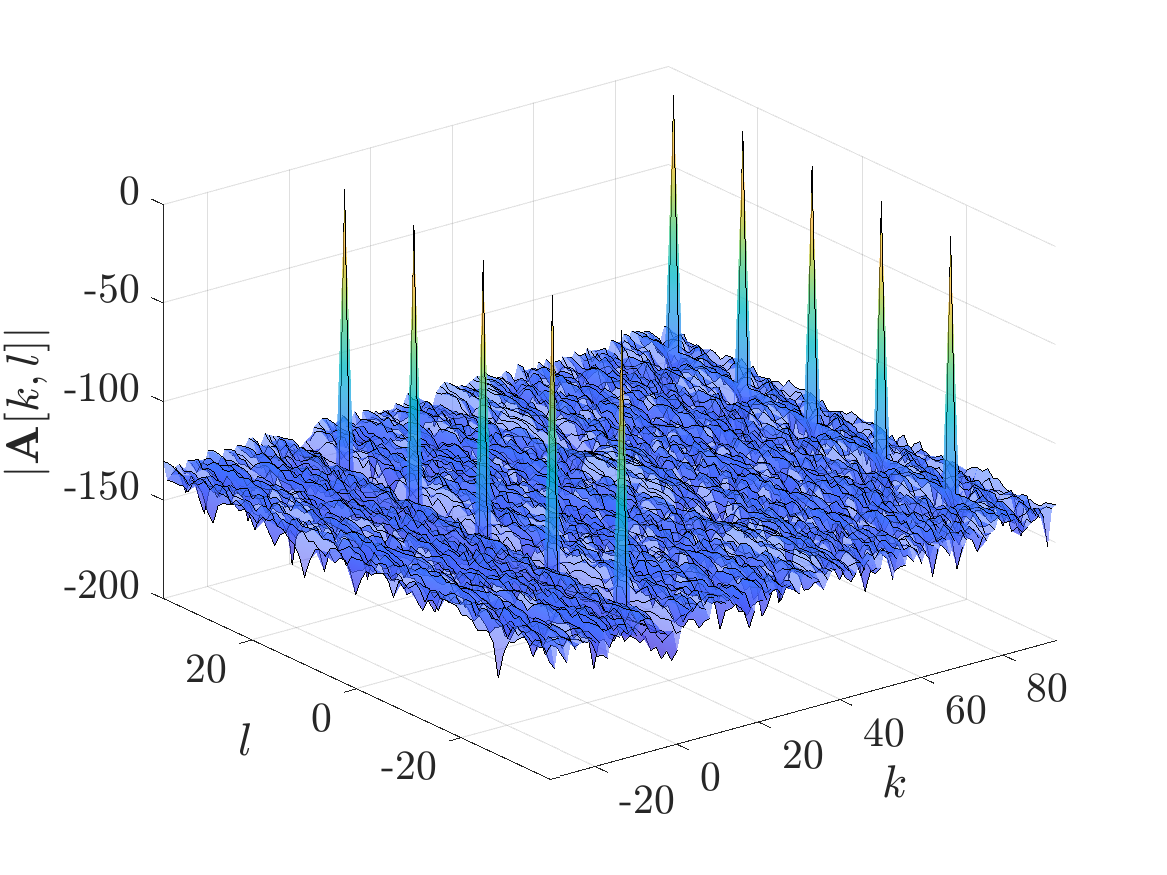}\label{fig:sa_proposed}}
    \hfill
    \caption{Absolute value of the TD phase coded and DD Zak-OTFS self-ambiguity (in log scale) for Zadoff-Chu sequences with root $u = 14$. The DD (TD) ambiguity has absolute value $1$ (close to $1$) on the line given by \eqref{eq:amb_final} with $\alpha = \beta = 7$. (a) TD AF of phase coded rectangular waveform as proposed in \cite{benedetto_phasecoded}, with parameters: $MN=1147$ chips of duration $1/B=1.0753 \mu$s each, and sampling rate $2B = 1.86$ MHz. (b) DD AF of the proposed method of modulating Zak-OTFS carrier waveforms with CAZAC sequences. The proposed DD AF has improved characteristics compared to TD phase coding, with no sidelobes at $(k,l)$ values outside the line~\eqref{eq:amb_final}.}
    \vspace{-4mm}
    \label{fig:sa}
\end{figure*}

In this Section, we derive the DD self- and cross-ambiguity properties of the array $\mathbf{X}[k,l]$ in~\eqref{eq:cazac_dd}. 


\subsection{DD Self-Ambiguity Function}
\label{subsec:selfamb_cazac}

We begin by deriving the DD self-ambiguity of the DD arrays in~\eqref{eq:cazac_dd}. We show that the DD self-ambiguity function has absolute value $1$ on the line $2\alpha k-l \equiv 0 \mod (MN)$, and has absolute value zero otherwise. For notational simplicity, we suppress the subscripts in the ambiguity function variable.

On substituting \eqref{eq:cazac_dd} in the self-ambiguity expression in Definition~\ref{def:ddamb}, we obtain:
\begin{align*}
    \mathbf{A}[k, l] &= \frac{1}{MN^2}\sum_{k'=0}^{M-1}\sum_{l'=0}^{N-1}\sum_{p=0}^{N-1}\sum_{q=0}^{N-1}e^{\frac{j2\pi}{MN}(\gamma+k'\beta)} \times \nonumber \\
    &\hspace{5mm}e^{\frac{j2\pi}{MN}\left(\alpha(k'+pM)^2 + pM(\beta-l')\right)}e^{-\frac{j2\pi}{MN}(\gamma+(k'-k)\beta)} \times \nonumber \\
    &\hspace{5mm}e^{-\frac{j2\pi}{MN}\left(\alpha(k'-k+qM)^2 + qM(\beta-l'+l)\right)} e^{-\frac{j2\pi}{MN}(k'-k)l} \nonumber \\
\end{align*}
\begin{align}
    &= \frac{1}{MN^2}e^{\frac{j2\pi}{MN}\left(lk+k\beta-\alpha k^2\right)}\sum_{p=0}^{N-1}e^{\frac{j2\pi}{MN}(\alpha p^2M^2+pM\beta)} \times \nonumber \\
    &\hspace{5mm}\sum_{q=0}^{N-1}e^{\frac{j2\pi}{MN}(-\alpha q^2M^2+2\alpha kqM-qM\beta - qMl)} \times \nonumber \\
    &\hspace{4.5mm}\sum_{k'=0}^{M-1}\hspace{-1.5mm}e^{\frac{j2\pi}{MN}(2\alpha k'pM + 2\alpha k'k - 2\alpha k'qM - lk')}\sum_{l'=0}^{N-1}\hspace{-1.5mm}e^{\frac{j2\pi}{N}(q-p)l'}.
    \label{eq:amb_1}
\end{align}

It follows from Identity~\ref{idty:sumrootsofunity} that the inner summation over $l'$ vanishes unless $p \equiv q \mod (N)$, when it takes value $N$. Since $0 \leq p, q \leq N-1$, $q-p \equiv 0 \mod (N)~\text{implies~} p = q$, and
\begin{align}
    \mathbf{A}[k, l] &= \frac{1}{MN}e^{\frac{j2\pi}{MN}\left(lk+k\beta-\alpha k^2\right)}\sum_{p=0}^{N-1}e^{\frac{j2\pi}{MN}(\alpha p^2M^2+pM\beta)} \times \nonumber \\
    &\hspace{5mm}e^{\frac{j2\pi}{MN}(-\alpha p^2M^2+2\alpha kpM-pM\beta - pMl)} \times \nonumber \\
    &\hspace{5mm}\sum_{k'=0}^{M-1}e^{\frac{j2\pi}{MN}(2\alpha k'pM + 2\alpha k'k - 2\alpha k'pM - kk')} \nonumber \\
    &= \frac{1}{MN}e^{\frac{j2\pi}{MN}\left(lk+k\beta-\alpha k^2\right)}\sum_{p=0}^{N-1}e^{\frac{j2\pi}{MN}(2\alpha kpM - pMl)} \times \nonumber \\
    &\hspace{5mm}\sum_{k'=0}^{M-1}e^{\frac{j2\pi}{MN}(2\alpha k'k -lk')}.
    \label{eq:amb_3}
\end{align}


Again, via Identity~\ref{idty:sumrootsofunity}, the inner summation over $k'$ takes value $M$ for all $\frac{2\alpha k-l}{N}\equiv 0 \mod (M)$. Since $M$ and $N$ are relatively prime, this implies that $2\alpha k-l\equiv 0 \mod (M)$.

Similarly, via Identity~\ref{idty:sumrootsofunity}, the sum over $p$ in \eqref{eq:amb_3} takes value $N$ for all $2\alpha k-l \equiv 0 \mod (N)$. Since $2\alpha k-l \equiv 0 \mod (M)$ and $2\alpha k-l \equiv 0 \mod (N)$ and $M$ and $N$ are co-prime, we have $2\alpha k-l \equiv 0 \mod (MN)$.

 

Thus, the DD self-ambiguity function in \eqref{eq:amb_3} reduces to:
\begin{align}
    \mathbf{A}[k, l] = \begin{cases}
        e^{\frac{j2\pi}{MN}(lk+k\beta-\alpha k^2)}, \text{ if } \ 2\alpha k-l \equiv 0 \hspace{-2mm} \mod (MN) \hspace{-3mm} \\
        0, \qquad\qquad\qquad \ \ \text{otherwise}
    \end{cases},
    \label{eq:amb_final}
\end{align}
with absolute value $1$ on the line $2\alpha k-l \equiv 0 \mod (MN)$, and absolute value $0$ otherwise. 

\subsection{DD Cross-Ambiguity Function}
\label{subsec:crossamb_cazac}

Next, we derive the DD cross-ambiguity of the DD sequences~\eqref{eq:cazac_dd}. We show that for any two sequences $\mathbf{x}[n],~\mathbf{y}[n]$ of the type~\eqref{eq:cazac} with $\alpha\neq\alpha',~\beta\neq\beta',~\gamma\neq\gamma'$ such that $(\alpha - \alpha')$ is co-prime to $MN$, the DD cross-ambiguity function has constant absolute value $\frac{1}{\sqrt{MN}}, \text{for all~} k,l$.

Direct substitution in the cross-ambiguity expression (Definition~\ref{def:ddamb}) gives: 
\begin{align*}
    \mathbf{A}_{\mathbf{X},\mathbf{Y}}[k,l] &= \frac{1}{MN^2}\sum_{k'=0}^{M-1}\sum_{l'=0}^{N-1}\sum_{p=0}^{N-1}\sum_{q=0}^{N-1}e^{\frac{j2\pi}{MN}(\gamma+k'\beta)} \times \nonumber \\
    &\hspace{5mm}e^{\frac{j2\pi}{MN}\left(\alpha(k'+pM)^2 + pM(\beta-l')\right)}e^{-\frac{j2\pi}{MN}(\gamma'+(k'-k)\beta')} \times \nonumber \\
\end{align*}
\begin{align}
    &\hspace{5mm}e^{-\frac{j2\pi}{MN}\left(\alpha'(k'-k+qM)^2 + qM(\beta'-l'+l)\right)} e^{-\frac{j2\pi}{MN}(k'-k)l} \nonumber \\
    &= \frac{1}{MN^2}\hspace{-1mm} \sum_{k'=0}^{M-1}\hspace{-0.75mm}\sum_{p=0}^{N-1}\hspace{-0.75mm}\sum_{q=0}^{N-1}\hspace{-1mm} e^{-\frac{j2\pi}{MN}(k'-k)l}\hspace{-1mm}\sum_{l'=0}^{N-1}\hspace{-1mm}e^{\frac{j2\pi}{N}(q-p)l'} \times \nonumber \\
    &\hspace{5mm}e^{\frac{j2\pi}{MN}(k'\beta-(k'-k)\beta')} e^{\frac{j2\pi}{MN}\left(\alpha(k'+pM)^2 + pM\beta\right)} \times \nonumber \\ 
    &\hspace{5mm}e^{-\frac{j2\pi}{MN}\left(\alpha'(k'-k+qM)^2 + qM(\beta'+l)\right)} e^{\frac{j2\pi}{MN}(\gamma-\gamma')}.
    \label{eq:crossamb_1}
\end{align}

Again, the inner summation over $l'$ vanishes unless $p=q$, when it takes value $N$. 
\begin{align}
    \label{eq:crossamb_2}
    &\mathbf{A}_{\mathbf{X},\mathbf{Y}}[k,l] = \frac{1}{MN} e^{\frac{j2\pi}{MN}\left(lk+k\beta'-\alpha' k^2\right)} e^{-\frac{j2\pi}{MN}\frac{(\beta-\beta'-l+2\alpha'k)^2}{4(\alpha-\alpha')}} \times \nonumber \\
    &e^{\frac{j2\pi}{MN}(\gamma-\gamma')} \sum_{k'=0}^{M-1}\sum_{p=0}^{N-1} e^{\frac{j2\pi}{MN}(\alpha-\alpha')\bigg(k'+pM+\frac{(\beta-\beta'-l+2\alpha'k)}{2(\alpha-\alpha')}\bigg)^2}.
\end{align}

To simplify the inner summation over $k',p$, or equivalently over the variable $t = k'+pM$, we utilize the following identity.

\begin{identity}[\cite{preamblepaper}]
    \label{idty:sumsquaredrootsofunity}
    For an odd integer $N$ and an integer $a$ co-prime to $N$,
    \begin{align}
        \sum_{n = 0}^{N-1} e^{\frac{j2\pi}{N} an^2} = C \sqrt{N},
    \end{align}
    where $C$ is a complex phase, i.e., $|C| = 1$.
\end{identity}

Applying Identity~\ref{idty:sumsquaredrootsofunity} to~\eqref{eq:crossamb_2}, we observe that the DD cross-ambiguity has constant absolute value $\frac{1}{\sqrt{MN}}, \text{for all~} k,l$. 


\section{Numerical Results}
\label{sec:implications}

\subsection{Implications for Radar}
\label{subsec:radar_impl}

CAZAC sequences are used to phase code radar waveforms~\cite{benedetto_phasecoded}. The characteristics of the ambiguity function depend on both the phase coding and the choice of waveform. Fig.~\ref{fig:sa} illustrates the importance of joint design. Fig.~\ref{fig:sa}\subref{fig:sa_benedetto} plots the TD AF for a ZC phase coded rectangular waveform as defined in~\cite{benedetto_phasecoded}. Fig.~\ref{fig:sa}\subref{fig:sa_proposed} plots the DD AF for ZC modulated Zak-OTFS carrier waveforms. While the DD (TD) ambiguity functions have absolute value $1$ ($\approx 1$) on the line given by~\eqref{eq:amb_final}, it is clear from the figure that TD phase coding results in sidelobes at $(k,l)$ values outside the line~\eqref{eq:amb_final}, thus limiting target detection. In contrast, the proposed method of modulating Zak-OTFS carriers has improved AF characteristics with no sidelobes at $(k,l)$ values outside the line~\eqref{eq:amb_final}. 


\subsection{Implications for Integrated Sensing \& Communication}
\label{subsec:isac_impl}
In all the simulations presented hereafter, we consider Zak-OTFS with $M=31, N=37$, Doppler period $\nu_p = 30$ KHz. We consider a $P=6$ path Vehicular-A (Veh-A) channel defined by 3GPP \cite{veh_a}, which has significant mobility and delay spread. The Doppler spread of each path $\nu_i = \nu_{\max}\cos(\theta_i), i=0, 1, \cdots, P-1$, where $\theta_i$ is uniformly distributed in the interval $[-\pi, \pi)$ and $\nu_{\max}$ is the maximum Doppler spread. Note that the delay and Doppler of each path is \textit{fractional} in nature. For pulse shaping, a Root Raised Cosine (RRC) filter with $\beta_\tau = \beta_\nu = 0.6$ is used.

\begin{figure}
   \centering
   \includegraphics[width=0.9\linewidth]{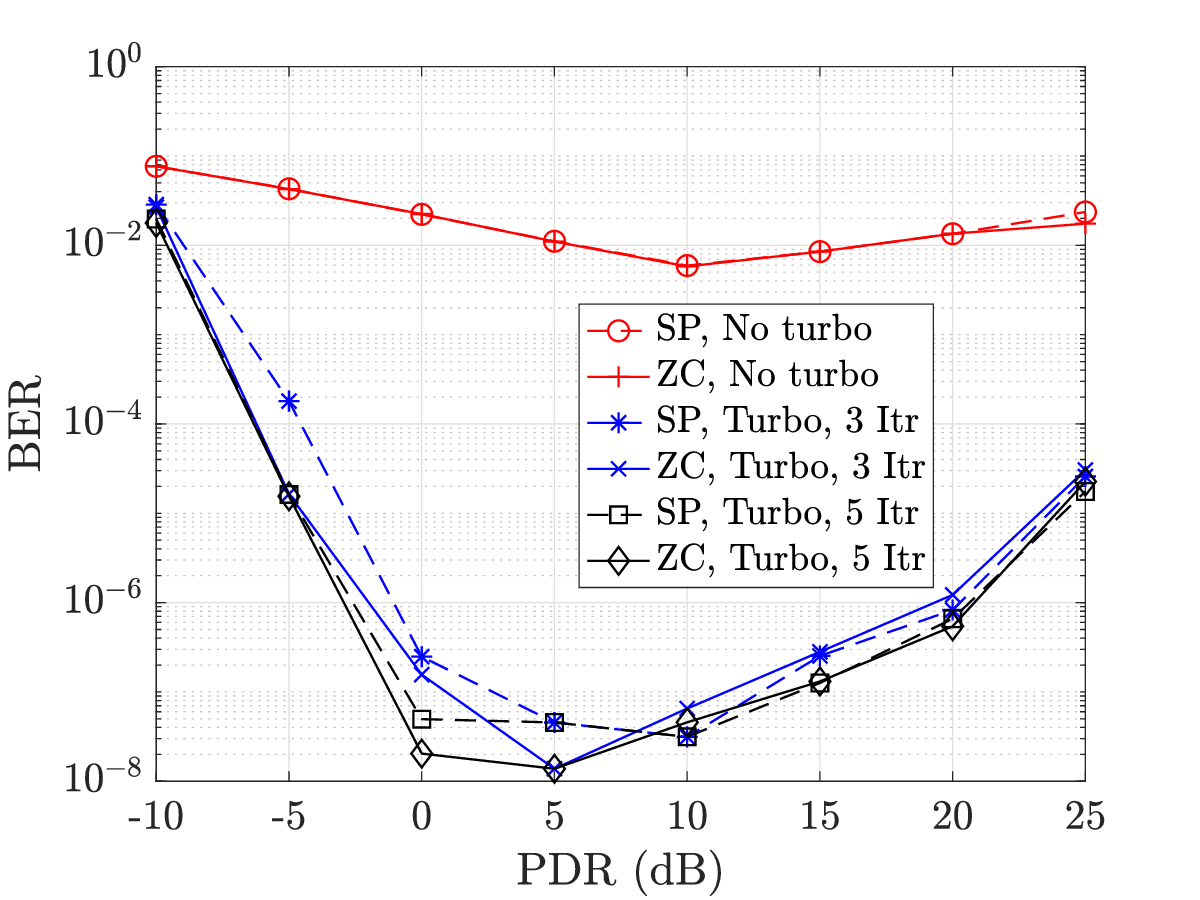}
   \caption{BER performance for uncoded 4-QAM as a function of increasing PDR. Superimposed frame is considered. ZC spread pilot with root $u=11$. Data SNR $\rho_d = 25$ dB, $\nu_{\max} = 6000$ Hz. Very significant improvement in BER performance from three to five turbo iterations.}
   \label{fig:turbo}
   \vspace{-3mm}
\end{figure}

Fig.~\ref{fig:turbo} shows the BER performance of uncoded 4-QAM as a function of increasing pilot to data ratio (PDR) \cite{Aug2024paper}. We use a Zadoff-Chu sequence with root $u=11$ in the DD domain as the spread pilot. The spread pilot is superimposed on the data. Since the data symbols and spread pilot are mutually unbiased, it is possible to sense the channel from the spread pilot and detect data from the data frame. To further improve the performance, we perform turbo iterations that alternate between channel sensing and data detection (see \cite{preamblepaper}[Sec.~V]). We note that the turbo iterations help achieve very good BER performance while maintaining high throughput.

\subsection{Implications for Uplink Preamble Detection}
\label{subsec:uplink_impl}

\begin{figure}[!t]
    \centering
    \includegraphics[width=0.9\linewidth]{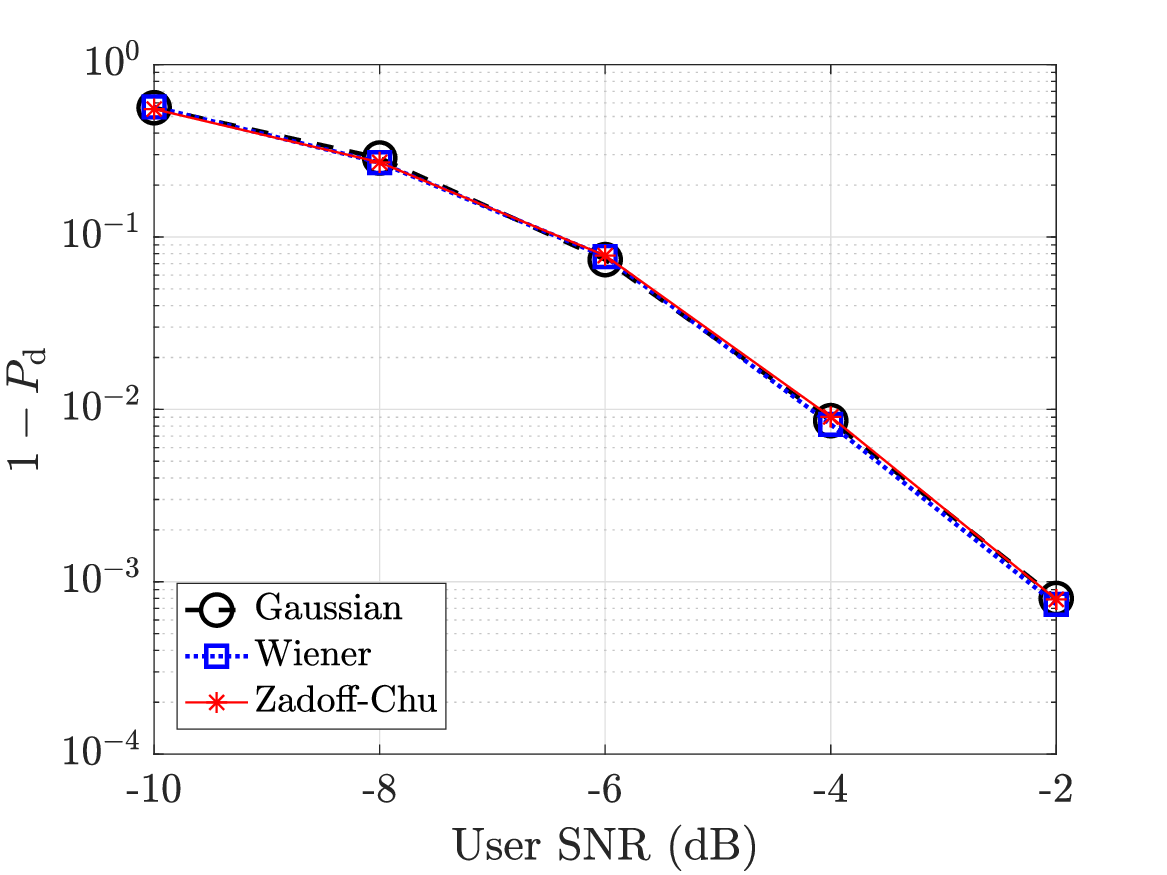}
    \caption{Probability of missed detection ($1-P_d$) as a function of user SNR for $K=5$ active users using different CAZAC sequences as preambles. The CAZAC sequence parameter $\alpha$ (and $\beta$ for Gaussian) was chosen to be co-prime to $MN$, $\gamma = 0$, and $\nu_{\max} = 815$ Hz. }
    \label{fig:ost_preamble}
    \vspace{-3mm}
\end{figure}

To demonstrate the benefits of our proposed method in simultaneous preamble detection for the 2-step RACH on the uplink, we consider preambles to be the DD CAZAC arrays defined in~\eqref{eq:cazac_dd}. For detection, we use the one-step thresholding (OST) algorithm, as described in~\cite{preamblepaper}. Fig.~\ref{fig:ost_preamble} shows that the probability of missed detection is similar regardless of which CAZAC sequence is used. The ability to simultaneously detect a modest number ($K=5$) of preambles enables grant-free multiple access in the presence of mobility and delay spread. 

\section{Conclusion}
\label{sec:conclusion}

Standard practice is to segregate sensing and communication, for example by introducing guard bands around sensing waveforms where no data is transmitted. Integration of sensing and communications then reduces to optimizing the balance between resources allocated to the two functions. We have proposed an alternative architecture based on coexistence rather than segregation. We have designed waveforms for sensing and communications that are mutually unbiased, so that interference between sensing and communications manifests as Gaussian noise. This makes it possible to optimize sensing and communications independently in the same radio resources. We have described how families of CAZAC sequences determine families of noise-like waveforms with excellent PAPR and ideal ambiguity properties. When used as pilots, we have shown that these waveforms can be superimposed on a full frame of data, so there is no need to sacrifice data rate to accommodate sensing. We have shown that the cross-ambiguity function between waveforms in the same family is flat, making it possible to simultaneously detect multiple waveforms in the presence of mobility and delay spread. This makes the waveforms ideal preambles supporting grant-free access on 6G wireless uplinks employing the 2-step RACH protocol in the 5GNR standard. 

\bibliographystyle{IEEEtran}
\bibliography{references}

\end{document}